\documentstyle[12pt]{article}
\newcommand{\be}{\begin{equation}}
\newcommand{\ee}{\end{equation}}
\newcommand{\bea}{\begin{eqnarray}}
\newcommand{\eea}{\end{eqnarray}}
\newcommand{\bdm}{\begin{displaymath}}
\newcommand{\edm}{\end{displaymath}}

\title{Large Squeezing Behavior of Cosmological \\ Entropy
Generation}
\author{M.\ Kruczenski \\
Departamento de F\'\i sica, TANDAR \\
Comisi\'on Nacional de Energ\'{\i}a At\'omica \\
Av. Libertador 8250, (1429) Buenos Aires, Argentina\\ \\
L.\ E.\ Oxman $\! ^*$ and M.\ Zaldarriaga \\
Departamento de F\'\i sica, Facultad de Ciencias
Exactas y Naturales \\
Universidad de Buenos Aires, Ciudad Universitaria \\
(1428) Buenos Aires, Argentina \\ \\
$^*$ Consejo Nacional de Investigaciones Cientif\'\i cas \\
y T\'ecnicas, Argentina}

\date{February 10, 1994}

\sf

\begin{document}
\maketitle
\newpage
\begin{abstract}

We consider the generation of entropy when particle pairs are
created at a cosmological level.
Making a reduction via the particle number basis, we compute the
classical limit for the entropy generation due to the evolution of
the matter field fluctuations (squeeze transformation), obtaining
that it is linear in the squeeze parameter for a general class of
initial states.

We also discuss the dependence of the generated entropy on the coarse
graining criteria. \\


\end{abstract}

\section{Introduction}

The cosmological particle production that occurs in early
universe models, because of the changing space-time metric, could be
a relevant source for entropy generation. In this respect, work has
been done introducing an entropy measure that increases as
the quantum field that represents these particles evolves.

If we approximate the metric by a given time-dependent background,
the evolution of the matter field fluctuations can be modelled;
while, as noted by L.\ P.\ Grishchuk and Y.\ V.\ Sidorov \cite{GrSi},
the corresponding Bogoliubov transformation that gives the relation
between the {\it in} and {\it out} creation and annhilation operators
can be expressed as a squeeze transformation.

The generation of entropy due to particle production can also be
studied in the formalism of squeezed states. The problem is that of
obtaining in an unambiguous way, a notion of entropy sensitive to the
unitary squeeze transformation $\hat{S}$ that the field undergoes.

In Ref. \cite{HP}, B.\ L.\ Hu and D.\ Pavon associated the entropy
generation with the increase of the mean particle number and the
loss of coherence of the initial state.

Alternatively, in Ref. \cite{Pro}, T.\ Prokopec considers the entropy
as that obtained by coarse graining the density matrix, that is, by
reducing the density matrix with respect to a given basis. In that
work the entropy change is considered, when the initial state is the
vacuum and the reduction is done via the occupation number basis or
the (over-complete) basis of coherent states. In both cases, the
classical limit (large squeezing limit) for entropy generation
coincides.

Another possibility, recently  suggested by M.\ Gasperini and M.\
Giovannini \cite{GG}, is to use the basis of eigenstates of $\hat{x}$,
the superfluctuant quadrature of the field. In the
$\hat{x}$-basis the calculation can be exactly done, the generated
entropy is linear in the squeeze parameter for all values of
squeezing (see Refs. \cite{BO} and \cite{GGLett}).

In general, we could think in coarse graining a given density matrix
$\rho$ with respect to an observable $\hat{A}$, having eigenfunctions
$|a\rangle$, and define the entropy as
\be
{\cal S}=-\sum_a \langle a|\rho |a \rangle
\ln \langle a|\rho |a\rangle
\label{ro}
\ee
Then, an important question that arises is to what extent the
generation of entropy due to the squeezing process depends on the
initial state and the coarse graining criteria.

In order to study these issues we will take the evolution of the
system in the Schr\"odinger picture. If one considers a pure state
$|\psi\rangle$, Eq. (\ref{ro}) reduces to \be
{\cal S}=-\sum_a |\langle a|\psi \rangle |^2
\ln |\langle a|\psi \rangle |^2
\ee
This is indeed the Shannon entropy for the probability distribution,
in the state $|\psi \rangle $, associated with the observable
$\hat{A}$. The intuitive meaning of this expression is that
$e^{\cal S}$ is the number of basis vectors which are ``appreciably
involved'' in the representation of $|\psi\rangle$ (``its
richness''), see Ref. \cite{qcaos}.

Therefore, given a coarse graining criteria, we are interested in the
way the richness of the initial state of the field changes, due to
the parametric squeezing (production of particle pairs).

In Ref. \cite{GrSi} each mode $\vec{k}$ of the field is independently
associated with a squeeze transformation. A better description is
given in Refs. \cite{Pro} and \cite{GG} where a two
mode transformation that mixes the modes $\vec{k}$ and
$-\vec{k}$ is considered, conserving the momentum during the
production of pairs.

For the sake of simplicity we will first consider the change of the
Shannon entropy under a one mode squeeze transformation
$|\psi\rangle \rightarrow \hat{S}|\psi\rangle$,
$\hat{S}=e^{-i\hat{G}}$, where
\be
\hat{G}=\frac{r}{2}i(a^{\dagger 2} -a^2)
\makebox[.5in]{,}
r>0
\label{sqop}
\ee
($a^{\dagger}$, $a$ are the usual creation and annhilation operators).

In Ref. \cite{BO} we studied the interplay between the quantum
fluctuations in the superfluctuant (resp. squeezed) quadrature and
the corresponding loss (resp. gain) of information.

In this article we will study the classical limit for
entropy generation showing that it is the same for a general class of
initial states (here the coarse graining is done via $\hat{N}$, the
particle number operator). Secondly, we will study the dependence of
the classical limit on coarse graining by considering some
particular examples.

In section \S 2 we review the eigenstates of $\hat{G}$ and their
properties which lead to a very simple description of the state
evolution. In section \S 3 we compute the leading order behavior (in
the squeezing parameter $r$) for $\hat{A}=\hat{N}$ considering a
general class of normalizable initial states (which include
particle number eigenstates). In section \S 4 we discuss
the dependence of the entropy generation on the coarse graining
criteria, and finally, in section \S 5 we compute the two mode case.

\section{The Eigenstates of the Squeeze Operator and the Propagation
Kernel}

In order to study the physical consequences of the squeezing
process it is convenient to use the wave function representation and
in particular a set of eigenfunctions of the squeeze generator
(\ref{sqop}). By defining the two canonically conjugate quadrature
operators:
\be
\hat{x}=\frac{1}{\sqrt 2}(a+a^{\dagger})
\makebox[.5in]{,}
\hat{p}=\frac{1}{\sqrt 2}i(a^{\dagger}-a)
\ee
we get a simple form for the squeeze generator $\hat G$, that
corresponds to a dilation transformation:
\be
\hat{G}=\frac{r}{2}(\hat{x}\hat{p}+\hat{p}\hat{x})
\makebox[1in]{,} [ \hat{x},\hat{p}]=i
\label{Ha}
\ee
The eigenfunctions of the dilation operator (\ref{Ha}) have been
introduced by C.\ G.\ Bollini and J.\ J.\ Giambiagi (see Ref.
\cite{BG}).
By choosing the following realization for $\hat{x}$ and $\hat{p}$
\begin{displaymath}
\hat{x}=x \makebox[1in]{,} \hat{p}=\frac{1}{i} \frac{d~}{dx}
\end{displaymath}
the eigenvalue equation $\hat G \psi=\mu r \psi$ reduces to
\begin{equation}
x\frac{d~}{dx}\psi=(i\mu -\frac{1}{2})\psi
\end{equation}
which leads to the solutions
\begin{equation}
\psi_+^{\mu} =\frac{1}{\sqrt{2\pi}}x_+^{i\mu -\frac{1}{2}}
\makebox[1in]{and}
\psi_-^{\mu} =\frac{1}{\sqrt{2\pi}}x_-^{i\mu -\frac{1}{2}}
\label{eigend}
\end{equation}
where
\begin{displaymath}
x_+^{\lambda}=\left\{ \begin{array}{ll}
x^{\lambda} &\mbox{if $x>0$} \\
0&\mbox{if $x<0$}
\end{array}  \right.  \makebox[1in]{and}
x_-^{\lambda}=\left\{ \begin{array}{ll}
0 &\mbox{if $x>0$} \\
|x|^{\lambda}&\mbox{if $x<0$}
\end{array}  \right.
\end{displaymath}
Note that the spectrum is continuous and extends from $-\infty$ to
$+\infty$.

The functions given in (\ref{eigend}) form a complete set and satisfy
$\delta$-function normalization:
\begin{equation}
\int_{-\infty}^{+\infty}d\mu
(\psi_+^{\mu}(x_1)\overline{\psi}_+^{\mu}(x_2)+
\psi_-^{\mu}(x_1)\overline{\psi}_-^{\mu}(x_2))
=\delta (x_1 -x_2)
\end{equation}
\begin{equation}
\langle \psi_{\pm}^{\mu_1}|
\psi_{\pm}^{\mu_2}\rangle =\delta (\mu_1 -\mu_2)
\makebox[.5in]{,}
\langle \psi_+^{\mu_1}|
\psi_-^{\mu_2}\rangle = 0
\end{equation}
In Ref. \cite{BO} we obtained the kernel associated with the squeezing
operator $\hat{S}=e^{-i\hat{G}}$, i.e., $K(x,x')=\langle
x|\hat{S}|x'\rangle$, which can be rewritten using the completness
relation:
\begin{equation}
K(x,x')=\frac{1}{2\pi}\int_{-\infty}^{+\infty}d\mu
\frac{1}{\sqrt{x' x}}(x_+^{i\mu} {x'_+}^{-i\mu}+
x_-^{i\mu} {x'_-}^{-i\mu}) e^{-i\mu r}
\label{propx}
\end{equation}
Therefore, if $x'$ and $x$ are greater than zero we obtain
\begin{eqnarray}
K(x,x')&=&\frac{1}{2\pi}\int_{-\infty}^{+\infty}d\mu
\frac{1}{\sqrt{x' x}}e^{i\mu (\ln x -\ln x' -r )}
\nonumber  \\
&=&\frac{1}{\sqrt{x' x}}\delta (\ln x -\ln x' -r )
\label{ppx}
\end{eqnarray}
If $x'$ and $x$ are both negative,
\begin{eqnarray}
K(x,x')&=&\frac{1}{\sqrt{x' x}}
\delta (\ln |x| -\ln |x'| -r )
\label{mmx}
\end{eqnarray}
and if $x'$ and $x$ have opposite signs, we see from (\ref{propx})
that $K(x,x')$ is zero.
These properties can be summarized by noting that the kernel
$K(x,x')$ is non zero only for $x=x' e^{r}$:
\begin{equation}
K(x,x')=e^{-\frac{r}{2}}\delta (x'- x e^{-r})
\label{Kx}
\end{equation}
Analogously, we get:
\be
K(p,p')=e^{\frac{r}{2}}\delta (p'- p e^{r})
\ee
This can be directly seen from the symmetry of our system under the
change $\hat{x}\rightarrow \hat{p}, \hat{p}\rightarrow -\hat{x},
r \rightarrow -r$ (cf. (\ref{Ha})).

{}From (\ref{Kx}) we see that under squeezing, an initial state
given by $\langle x|\psi\rangle=\psi(x)$ evolves into
\begin{eqnarray}
\langle x|\hat{S}|\psi\rangle &=&\int_{-\infty}^{+\infty}dx'
K(x|x')\psi (x')
\nonumber  \\
&=&e^{-\frac{r}{2}}\int_{-\infty}^{+\infty}dx' \delta
(x' -x e^{-r })\psi(x')
\nonumber  \\
&=&e^{-\frac{r}{2}}\psi (xe^{-r})
\label{psit}
\end{eqnarray}
Similarly,
\be
\langle p|\hat{S}|\psi\rangle=e^{\frac{r}{2}}\varphi (p e^{r})
\label{fit}
\ee
where $\varphi (p)=\langle p|\psi\rangle$.

It is straightforward to verify that the normalization is preserved
during the evolution.

We can see that, for $r>0$, $\hat{x}$ (resp. $\hat{p}$) is the
superfluctuant (resp. squeezed)
quadrature: using (\ref{psit}) and (\ref{fit}) we have
for the dispersions in $\hat{x}$ and $\hat{p}$,
\be
\sigma_r(\hat{x})=e^{r}\sigma_0(\hat{x})
\makebox[.5in]{,}
\sigma_r(\hat{p})=e^{-r}\sigma_0(\hat{p})
\ee
Note also that the eigenfunctions (\ref{eigend}) are the
analog to plane waves but instead of being invariant (up to a phase)
under translations, they are invariant under dilations.

\section{Entropy Generation and the Initial State}

We will consider here the entropy generation which is obtained
by coarse graining the density matrix. To do so we
have to choose an observable $\hat{A}$ with eigenfunctions
$|a\rangle$ and write the density matrix as
\be
\rho =\sum_a \sum_{a'} |a\rangle \langle a| \rho |a'\rangle
\langle a'|
\ee
Then, this density matrix is reduced by setting to zero the
off-diagonal terms to get:
\be
\rho_{red} =\sum_a |a\rangle \langle a| \rho |a\rangle
\langle a|
\label{red}
\ee
and the associated expression for the entropy is
\bea
{\cal S}&=&-Tr\, \rho_{red} \ln \rho_{red} \nonumber \\
&=&-\sum_a \langle a| \rho |a\rangle \ln
\langle a| \rho |a\rangle
\label{entro}
\eea
In contrast with the entropy computed with the full density matrix,
which is unaltered by squeezing, the entropy (\ref{entro}) is
sensitive to this evolution and depends on the reduction scheme.
A natural scheme is that associated with the occupation number
basis, for which we will study the large squeezing regime
($r>>1$) of the entropy generation.

We will consider an initial pure state $|\psi \rangle$ which is
supposed to be normalizable. The corresponding density matrix $\rho
=|\psi\rangle \langle \psi|$ leads to a Shannon entropy with respect
to the particle number given by
\be
{\cal S}=-\sum_n \left|\langle n|\psi (r) \rangle \right|^2
\ln \left|\langle n|\psi (r) \rangle \right|^2
\label{Sr}
\ee
where $|\psi (r)\rangle =\hat{S}|\psi\rangle$.
Now, in order to obtain the large squeezing behavior of (\ref{Sr}),
we first note that the amplitudes $\langle n|\psi (r) \rangle$ go to
zero in that limit, for in the $\hat{x}$-representation we have
(cf. (\ref{psit})):
\bea
|\langle n|\psi (r) \rangle|&=&e^{-\frac{r}{2}}|
\int_{-\infty}^{+\infty}dx \, \psi_n(x)
\psi (e^{-r}x)|\nonumber \\
&\leq &e^{-\frac{r}{2}}M \int_{-\infty}^{+\infty}
dx\, \frac{|H_n(x)| e^{-\frac{x^2}{2}}}{\sqrt{2^n n!}\pi^{\frac{1}{4}}}
\label{npsic}
\eea
where $M$ is an upper bound on the values of $|\psi (x)|$,
$\psi_n(x)$ are the eigenfunctions of the harmonic oscillator and
$H_n(x)$ are the Hermite polynomials. As the integral in
(\ref{npsic}) is convergent, the overlapping $\langle n|\psi (r)
\rangle$ goes to zero at least exponentially for large $r$
(and each term in (\ref{Sr}) goes to zero at least as $re^{-r}$).

Then, for large squeezing, the leading order of (\ref{Sr}) comes
from the infinite sum, and we can compute it by summing from a given
$n_0$:
\be
{\cal S}\sim -\sum_{n=n_0}^{\infty}
|\langle n|\psi (r) \rangle |^2
\ln |\langle n|\psi (r) \rangle |^2
\makebox[.5in]{,} r>>1
\label{den0}
\ee
Now, in order to make the computation of the asymptotic behavior
easier we will consider the states in the $\hat{p}$-representation,
the quadrature which is being squeezed, and we will suppose that
$\varphi (p)$ is an even function (a similar reasoning applies for
$\varphi (p)$ odd). From Eq. (\ref{fit}) we have
\bea
\langle 2k|\psi (r) \rangle &=&\int_{-\infty}^{+\infty} dp\,
\varphi_{2k}(p) e^{\frac{r}{2}}
\varphi (e^r p)\nonumber \\
&=&2e^{\frac{r}{2}}
\left( \int_{0}^{\delta}+\int_{\delta}^{+\infty}\right)
dp\, \varphi_{2k}(p) \varphi (e^r p)
\label{amp1}
\eea
where $\delta$ is a fixed number. In the first integral we can use
(see \cite{Abram})
\be
\varphi_{2k}(p)= \frac{1}{\sqrt{\pi}k^{\frac{1}{4}}}
\cos (2p\sqrt{k+^1\!\!/_4} )(1+{\cal O}(\frac{1}{\sqrt{k}}))
\label{approx}
\ee
which is valid for large values of $k$ (and $p$ bounded). Therefore,
\bea
&&\langle 2k|\psi (r) \rangle
=\frac{2e^{\frac{r}{2}}}{\sqrt{\pi} k^{\frac{1}{4}}}
\left( \int_{0}^{+\infty}-\int_{\delta}^{+\infty}\right)
dp\, \cos (2p\sqrt{k+^1\!\!/_4})
(1+{\cal O}(\frac{1}{\sqrt{k}}))\varphi (e^r p)\nonumber \\
&&+2e^{\frac{r}{2}}
\int_{\delta}^{+\infty} dp\, \varphi_{2k}(p) \varphi (e^r p)
\eea
In order to set an upper bound on the integrals over the interval
$(\delta,\infty)$ we demand that the associated wave function
$\varphi (p)$ satisfy, for large $p$, $|\varphi (p)|\leq const.\,
p^{-2}$ and $|\frac{d\varphi (p)}{dp}| \leq const.\, p^{-3}$.
In particular, this condition is fulfilled  by the eigenstates of the
particle number operator. We end up with an expression of the form
\be
\langle 2k |\psi (r) \rangle = \frac{2 e^{\frac{r}{2}}}{\sqrt{\pi}
k^{\frac{1}{4}}} \int_{0}^{\infty} dp\, \cos (2 p \sqrt{k}) \left(
1+{\cal O}(\frac{1}{\sqrt{k}})\right) \varphi (e^r p) + {\cal O}\left(
\frac{e^{-r}}{\sqrt{k}} \right)
\ee
where we have replaced $k+1/4$ by $k$ within the same order of
approximation considered. Finally the change of variable
$p\rightarrow e^{-r} p$ leads to
\be
\langle 2k |\psi (r) \rangle = f(k e^{-2r}) e^{-r} + g(k e^{-2r})
{\cal O}(e^{-2r})
\label{disp}
\ee
with
\be
f(k)=2 \int_{0}^{\infty}dp\,
\frac{\cos(2p\sqrt{k})}{\sqrt{\pi}k^{\frac{1}{4}}} \varphi (p)
\label{fdek}
\ee
and $g(k)$ being a function which for large $k$ vanishes faster than
$1/\sqrt{k}$.

Now, replacing the sum by the integral, we can compute the leading
order behavior of (\ref{den0}):
\be
{\cal S}\sim -\int_{k_0}^{\infty}dk\,
|\langle 2k|\psi (r) \rangle |^2
\ln |\langle 2k|\psi (r) \rangle |^2
\ee
Choosing $k_0$ so as to make valid the assumption (\ref{approx}), and
using (\ref{disp}), we see that
\bea
{\cal S}&\sim &-\int_{k_0}^{\infty}dk\, e^{-2r}
f^2(ke^{-2r})
(-2r+\ln f^2(ke^{-2r}))\nonumber \\
&\sim & 2r \int_{k_0e^{-2r}}^{\infty}dk\, f^2(k)
-\int_{k_0e^{-2r}}^{\infty}dk\, f^2(k)\ln f^2(k)
\label{asympS}
\eea
as $k_0$ is independent of $r$ we can integrate the last expression
from zero (in the large squeezing regime). From (\ref{fdek}) we
see that
\bea
\int_0^{\infty}dk\, f^2(k)&=&\int_{-\infty}^{+\infty}dp'\, dp\,
\frac{1}{\pi}\int_0^{\infty}\frac{dk}{k^{\frac{1}{2}}}
\cos \left( 2\sqrt{k}p\right)
\cos \left( 2\sqrt{k}p'\right)
\bar{\varphi} (p')\varphi (p)\nonumber \\
&=&\int_{-\infty}^{+\infty}dp\,\bar{\varphi} (p)\varphi (p)~=~1
\label{norm}
\eea
The second term in (\ref{asympS}) is an approximation to the
series for the initial entropy. Although this approximation is not so
good, the difference with the exact series is $r$-independent and
does not affect the leading order behavior of the generated entropy.
Then, using (\ref{norm}), we finally get:
\be
\Delta {\cal S}={\cal S}-{\cal S}_0\sim 2r
\makebox[.5in]{,}
r>>1
\label{asymp1}
\ee
It is straightforward to generalize this result for a general
initial mixture of normalizable states $|b\rangle$:
\be
\rho = \sum_{b,b'} p_{bb'} |b\rangle \langle b'|
\makebox[.5in]{.}
Tr\, \rho =1
\label{rob1}
\ee
Following essentially the same steps that led from (\ref{Sr}) to
(\ref{asymp1}), for large values of squeezing, we also get
\bea
&&{\cal S}=-\sum_{b,b'} p_{bb'} \sum_k
\langle 2k |b(r) \rangle \langle b'(r) |2k\rangle
\ln \left( \sum_{b,b'}p_{bb'}
\langle 2k |b(r)\rangle \langle b'(r)|2k \rangle
\right) \nonumber \\
&&\sim 2r
\eea

\section{Entropy Generation and the Coarse Graining Criteria}

In the previous section we obtained that the generated entropy, when
coarse graining with respect to $\hat{N}$, goes like $\Delta {\cal
S}\sim 2r$, in the large squeezing regime. In general, if one
considers another coarse graining basis, a different
asymptotic behavior could be expected. For instance, if we take the
singular case of
the basis of the dilation operator eigenstates (cf. (\ref{eigend})),
the associated entropy does not change under a squeeze
transformation. This is precisely because these
functions are stationary under squeezing. Now, let us take as an
example the set of Laguerre functions
\be
\chi_{n}(x)=\frac{1}{\sqrt{2}}e^{-\frac{|x|}{2}}L_n (|x|)
\ee
which are similar (in the sense of their localization properties)
to the harmonic oscillator case, and let us consider the entropy
change under a squeeze transformation of the state
\be
\psi (x)=\frac{1}{\sqrt{2}} e^{-\frac{|x|}{2}}
\label{inst}
\ee
Using (see \cite{TRusa})
\be
\int_0^{+\infty}dx\, e^{-ax} L_n(x)=\frac{(a-1)^n}{a^{n+1}}
\makebox[.5in]{,}
\Re (a) > 0
\ee
we have
\be
\langle \chi_{n}|\psi (r)\rangle
=(-1)^n \frac{\tanh ^n (\frac{r}{2})}{\cosh (\frac{r}{2})}
\ee
and for the associated entropy we get
\bea
{\cal S}&=&-\sum_n|\langle \chi_{n}|\psi (r)\rangle|^2 \ln
|\langle \chi_{n}|\psi (r)\rangle|^2 \nonumber \\
&=& \ln \left(\cosh^2 (\frac{r}{2})\right) +
\sinh^2 (\frac{r}{2})\ln \tanh ^2 (\frac{r}{2})\nonumber \\
&\sim & r
\label{asymp2}
\eea
then, the leading order behavior of the generated entropy
gives $\Delta {\cal S}\sim r$.

There is however a factor of two with respect to the reduction done
via the $\hat{N}$-basis. This can be qualitatively understood by
using the intuitive meaning of the Shannon entropy of a state with
respect to a given basis \cite{qcaos}: $e^{\cal S}$ is the number of
basis vectors which are ``appreciably involved'' in the
representation of  $|\psi\rangle$. As the squeezing developes, the
width of the wave function in the $\hat{x}$-representation grows as
$e^r$. In the case of the $\hat{N}$-basis, in
order to represent $|\psi (r)\rangle$, the $|n\rangle$ which are
significant, can be estimated as those having $\sqrt{\langle
x^2\rangle_n}$ less than
the width of the wave function we are considering. Taking into
account that
for the harmonic oscillator $\sqrt{\langle x^2\rangle_n} \sim
n^{\frac{1}{2}}$, the number $n$ of basis states
that participate in the representation of $|\psi (r)\rangle$ verifies
$n^{\frac{1}{2}} \sim \sigma_0 e^r$ ($\sigma_0$ is the width of
$\psi (x)$) which gives ${\cal S}\sim \ln n \sim 2 r$
(cf. (\ref{asymp1})). On the
the other hand, in the case of the Laguerre functions, we have
$\sqrt{\langle x^2\rangle_n} \sim \sqrt{6} n$, so applying the
previous argument we obtain ${\cal S}\sim \ln n \sim r$ (cf.
(\ref{asymp2})).

We will now consider a coarse graining via the superfluctuant
quadrature $\hat{x}$. This is motivated by the approximated
eigenfunctions used in Eq. (\ref{disp}) which
are plane waves (Dirac $\delta$'s in the $\hat{x}$ representation).
This alternative basis was recently suggested by
M.\ Gasperini and M.\ Giovannini. In this case, the
calculation can be exactly done and the generated entropy turns out
to be linear for all values of squeezing (see Refs. \cite{BO} and
\cite{GGLett}). In Ref. \cite{GGLett}, this property was shown in the
case where the inital density matrix is diagonal in the occupation
number basis. From the state transformation given by Eq.
(\ref{psit}), regardless the initial density matrix, the reduction
done via the $\hat{x}$ operator leads to a linear generation of
entropy under
a squeeze transformation. If we make the reduction of (\ref{rob1})
using $\hat{A}=\hat{x}$ in equation (\ref{red}), we obtain:
\be
{\cal S}=- \sum_{b,b'} p_{bb'}
\int dx \langle x|b(r)\rangle  \langle b'(r)|x\rangle
\ln \left( \sum_{b,b'} p_{bb'} \langle x|b(r)\rangle
\langle b'(r)| x\rangle
\right)
\ee
where we have made explicit the squeezing of the initial state.
Now, using (\ref{psit}) we see that
\be
\langle x|
b(r)\rangle  \langle b'(r)| x\rangle
=e^{-r} \varphi_b (e^{-r}x)\bar{\varphi}_{b'} (e^{-r}x)
\ee
where $\varphi_b (x)$ is the wave function $\langle x|b\rangle$, so
the expression for the entropy is:
\bea
\lefteqn{{\cal S}=r\int dx \langle x|\hat{S} \rho \hat{S}^{\dagger}
| x\rangle} \nonumber \\
&&-\int dx e^{-r} \sum_{b,b'} p_{bb'}
\varphi_b (e^{-r}x)\bar{\varphi}_{b'} (e^{-r}x)
\ln \sum_{b,b'} p_{bb'}
\varphi_b (e^{-r}x)\bar{\varphi}_{b'} (e^{-r}x)
\eea
The second term is just the initial entropy as one can see by changing
variables $e^{-r}x \rightarrow x$; while the first term is
$r\, Tr\, \hat{S} \rho \hat{S}^{\dagger}=r\, Tr\, \rho =r$
so we get:
\be
\Delta {\cal S} = r
\ee
This coincides with what we would have expected from our qualitative
argument: the width of the state grows like $e^r$, so the ``number''
of $\delta$'s involved grows as $e^r$, and the entropy goes like $r$.

Finally, a reduction via the basis of coherent states can be considered.
Starting from the vacuum, the leading order behavior can be
calculated, giving $\Delta {\cal S}\sim r$. We can understand this
behavior by noting that the localization properties of the
over-complete basis of
coherent states are similar to those of the Dirac delta's: the width
is the same for every coherent state and they are uniformly
distributed on the real axis.

\section{The Two Mode Squeeze Operator}

In Ref. \cite{GrSi} each mode $\vec{k}$ of the field is independently
associated with a squeeze transformation. A better description is
given in Refs. \cite{Pro} and \cite{GG} where it is considered a two
mode transformation that mixes the modes $\vec{k}$ and $-\vec{k}$,
conserving the momentum during the production of pairs.

The evolution operator can be expressed as (see \cite{AFJP})
\be
\hat{U}=\hat{R}\, \hat{S}
\label{evol}
\ee
where
\bea
\hat{R}&=& e^{i\int_0^t dt'\, H_0(t')}
\makebox[.5in]{,}
H_0=\sum_{k,k_x>0}\Omega_k(t)(a_k^{\dagger}a_k +
a_{-k}^{\dagger}a_{-k}+1) \label{Hfree} \\
\hat{S}&=&e^{i\hat{G}}
\makebox[.5in]{,}
\hat{G}=\sum_{k,k_x>0} ir_k(t)\left(
e^{-2i\varphi_k (t)} a_k a_{-k}-
e^{2i\varphi_k (t)} a_k^{\dagger} a_{-k}^{\dagger}\right)
\label{Hint}
\eea
By defining new operators $b_k$ and $c_k$ ($k_x >0$) according to:
\be
a_k =\frac{1}{\sqrt{2}} (b_k -ic_k)e^{-ivarphi_k}
\makebox[.5in]{,}
a_{-k} =\frac{1}{\sqrt{2}} (b_k +ic_k)e^{-i\varphi_k}
\label{redef}
\ee
which satisfy the canonical commutation relations, $\hat{S}$ can be
expressed in terms of two single mode squeeze transformations of the
form given in (\ref{sqop}):
\be
\hat{S}=\prod_{k,k_x >0}
\exp \left( \frac{r_k}{2} (b_k^{\dagger 2}- b_k^2)\right)
\exp \left( \frac{r_k}{2} (c_k^{\dagger 2}- c_k^2)\right)
\label{sqk}
\ee

Now, we will consider the generation of entropy under the evolution
(\ref{evol}) (in each pair of modes $\vec{k}$ and $-\vec{k}$). For
simplicity, we will first consider as coarse graining basis:
$|n_b,n_c\rangle$ (the eigenstates of $\hat{N}_b=b^{\dagger}_k b_k$
and $\hat{N}_c=c^{\dagger}_k c_k$) and then the more natural
one, $|n_k,n_{-k}\rangle$ (the eigenstates of
$\hat{N}_k=a^{\dagger}_k a_k$ and
$\hat{N}_{-k}=a^{\dagger}_{-k}a_{-k}$).

In the first case we have that the entropy at time $t$ is given by
\be
{\cal S}=-\sum_{n_b,n_c}
\left|\langle n_b,n_c|\hat{U}_k(t) |\psi \rangle \right|^2
\ln \left|\langle n_b,n_c|\hat{U}_k(t) |\psi \rangle \right|^2
\label{Srab}
\ee
where $\hat{U}_k(t)$ is the part of (\ref{evol}) that contains the
modes $\vec{k}$ and $-\vec{k}$. From (\ref{Hfree}) and (\ref{redef})
we see that the corresponding free part of the hamiltonian takes the
form $\Omega_k(t)(b^{\dagger}_k b_k +c^{\dagger}_k c_k +1)$ which
gives an irrelevant phase factor in the amplitudes of (\ref{Srab}).

Now, using an approximation similar to (\ref{disp}), but for two
single mode squezze operators, we obtain
\bea
&&\langle n_b,n_c|\hat{S}_k(t)
|\psi \rangle \approx e^{-2r} f(k_b e^{-2r},k_c e^{-2r})\nonumber \\
&&f(k_b ,k_c)= \int_{-\infty}^{\infty}\int_{-\infty}^{\infty}
dp_b\, dp_c\,
\frac{\cos(2p_b\sqrt{k_b})}{\sqrt{\pi}k_b^{\frac{1}{4}}}
\frac{\cos(2p_c\sqrt{k_c})}{\sqrt{\pi}k_c^{\frac{1}{4}}}
\varphi (p_b, p_c)
\label{tmapprox}
\eea
where $n_b=2k_b$, $n_c=2k_c$, $\hat{S}_k(t)$ is the
factor appearing in (\ref{sqk}) and we have suppossed, as we did in
(\ref{disp}), that the initial state is even. From (\ref{tmapprox}),
we finally get $\Delta {\cal S}(t)\sim 4 r_k(t)$.

The second case is more interesting because it corresponds to the
occupation number basis, which has a direct physical significance.
Let us consider for example an initial state having $n_0$ particles
with momentum $\vec{k}$ and $n_0$ particles with momentum $-\vec{k}$
(the total momentum is zero). Again, the free part of (\ref{evol})
plays no role and the entropy at time $t$ is
\be
{\cal S}=-\sum_{n_k ,n_{-k}}
\left|\langle n_k,n_{-k}|\hat{S}_k(t) |n_0,n_0 \rangle \right|^2
\ln \left|\langle n_k,n_{-k}|\hat{S}_k(t) |n_0,n_0 \rangle \right|^2
\label{Srkmk}
\ee
Taking into account that the squeeze operator $\hat{S}_k(t)$ creates
and annihilates pairs of particles in the modes $\vec{k}$ and
$-\vec{k}$ (cf. (\ref{Hint})), we have that the surviving terms in
(\ref{Srkmk}) are those having $n_k=n_{-k}=n$:
\be
{\cal S}=-\sum_n
\left|\langle n,n|\hat{S}_k(t) |n_0,n_0 \rangle \right|^2
\ln \left|\langle n,n|\hat{S}_k(t) |n_0,n_0 \rangle \right|^2
\label{Srkmkd}
\ee
We can change $|n,n\rangle$ to the basis $|n_a,n_b\rangle$:
\be
|n,n\rangle =\frac{(b_k^{\dagger 2}
+c_k^{\dagger 2})^n}{n!2^n} |0,0 \rangle =\frac{1}{2^n}\sum_{j=0}^{n}
\frac{\sqrt{(2n-2j)!(2j)!}}{(n-j)!j!}|2n-2j, 2j\rangle'
\label{chba}
\ee
where the prime in the second member denotes that the ket is given in
the basis $|n_b,n_c\rangle$.

Using Stirling's formula  we can approximate the factorials in
(\ref{chba}) as
\be
\frac{\sqrt{(2n-2j)!\,(2j)!}}{2^n(n-j)!\, j!} \approx
\frac{1}{\sqrt{\pi}[(n-j)j]^{\frac{1}{4}}}
\label{modint2}
\ee
moreover, for $n$ large we shall replace the sum by an integral.
Note however that in spite of the approximation we still have the
normalization condition:
\be
\int_0^n dj\,
\frac{1}{\pi[(n-j)j]^{\frac{1}{2}}}=\frac{1}{\pi}
B(\frac{1}{2},\frac{1}{2})=1
\makebox[.3in]{,}
B(\alpha,\beta)=
\frac{\Gamma (\alpha)\Gamma (\beta)}{\Gamma (\alpha +\beta)}
\ee
Now, for large squeezing, we obtain from (\ref{tmapprox})
\bea
\langle n_0,n_0|\hat{S}^{\dagger}_k(t)|n,n\rangle
&\approx &e^{-2r}\int_0^n dj\,
\frac{1}{\sqrt{\pi}[(n-j)j]^{\frac{1}{4}}}
f((n-j) e^{-2r},j e^{-2r})\nonumber \\
&=&\frac{e^{-r}}{\sqrt{\pi}}\int_0^{ne^{-2r}} dj\,
\frac{1}{[(ne^{-2r}-j)j]^{\frac{1}{4}}}
f(ne^{-2r}-j,j)\nonumber \\
&=& e^{-r} \tilde{f}(ne^{-2r})
\label{ultapp}
\eea
Approximating the sum in (\ref{Srkmkd}) by the integral and using
(\ref{ultapp}) as well as the result (see Appendix)
$\int^{\infty}_{0}dn\, \tilde{f}(n)^2 =1 $ (similar to (\ref{norm}))
we obtain the leading order behavior for the generated
entropy in the occupation number basis:
\be
\Delta {\cal S}(t)\sim 2r_k(t)
\label{entgrowth}
\ee

\section{Discussion}

In this paper we studied the dependence of the entropy
generation on the initial state and the coarse graining criteria,
relating these issues to the Shannon entropy; i.e., the change of the
richness of a state in a given basis as the squeezing developes. For
this purpose we considered the eigenstates of the squeeze operator
which enabled us to describe the evolution as a dilation
transformation of the wave function.

Making a reduction via the particle number basis, we computed
the classical limit for the entropy generation under a one mode
squeeze transformation, obtaining that for a general class of initial
states it is linear in the squeeze parameter. This is also the case
when the reduction is done via the superfluctuant quadrature of the
field (here, the linear behavior is valid for any squeezing).

A deep understanding of entropy generation due to the creation of
particle pairs will come from the knowledge of the physical mechanism
that singles out a coarse graining basis. We note however that for a
wide class of basis, the associated entropy will increase as the
system evolves. This generation of entropy can be understood if we
recall that the
Shannon entropy gives the richness of a state with respect to a given
basis. In the superfluctuant quadrature representation the wave
function of the initial state flattens (cf. (\ref{psit})); and
for a basis sharing the localization properties of the occupation
number basis or that of Dirac's $\delta$-functions we will have that
at each stage more basis vectors are necessary to describe the state.

Then, we studied the entropy generation (in each pair of modes
$\vec{k}$ and $-\vec{k}$) when the matter field undergoes a two mode
squeeze evolution, that is, when particle pairs are created or
annhilated from the initial state, conserving momentum. Here, we
considered the reduction via two different basis: the
occupation number basis for the modes $b_k$ and $c_k$ that factorize
the two mode squeeze transformation, and the occupation number basis
for particles with momentum $\vec{k}$ and $-\vec{k}$. In the first
case the classical behavior of the generated entropy is
${\cal S}\sim 4r_k (t)$ (twice the value for the one mode case) while
in the second case, starting from a state with definite particle
number and zero momentum, it is ${\cal S}\sim 2r_k (t)$. This
calculation generalizes the result given in Ref. \cite{Pro} where the
initial state is the vacuum.

Then, in the cases we have considered, we can see that the initial
state does not leave its trace, when we look at the classical
behavior of the generated entropy. This behavior turns out to be
linear in the squeeze parameter with a proportionality factor that
depends on the coarse graining criteria.

\section*{Acknowledgments}

We would like to thank D.\ Harari and E.\ Calzetta for
usefull discussions on related matters.

\newpage
\section*{Appendix}

Here we shall show the result used in (\ref{entgrowth}):
\be
\int^{\infty}_{0}dn\, \tilde{f}(n)^2 = 1
\label{norm2}
\ee

In order to obtain an expression for $\tilde{f}(n)$ we should find
the equivalent of equations (\ref{approx}) and (\ref{fdek}) for the
two mode case.

First, using (\ref{modint2}) and (\ref{approx}) we have
\bea
\langle p_b p_c |nn\rangle &\approx&
\frac{1}{\sqrt{\pi}}\int_{0}^{n}dj\, \frac{1}{[(n-j)j]^{\frac{1}{4}}}
\langle p_bp_c|2n-2j,2j\rangle' \nonumber\\
&\approx &\frac{1}{\pi^{\frac{3}{2}}}\int_{0}^{n}dj\,
\frac{1}{[(n-j)j]^{\frac{1}{2}}} \cos(2p_b\sqrt{n-j})\cos(2p_c\sqrt{j})
\nonumber\\
&=&\frac{1}{\sqrt{\pi}} J_0\left( 2\sqrt{n(p_b^2+p_c^2)}\right)
\eea
$J_0$ is a Bessel function.

Now we can compute $\tilde{f}(n)$,
\bea
\tilde{f}(n)&=&\int dp_b\, dp_c\, \langle n_0n_0|p_bp_c\rangle
\frac{1}{\sqrt{\pi}}J_0\left( 2\sqrt{n(p_b^2+p_c^2)}\right) \\
&=&\frac{1}{2^{n_0}}\sum_{j=0}^{n_0}\frac{\sqrt{(2n_0-2j)!
(2j)!}}{\sqrt{\pi}(n_0-j)!j!}
\int dp_b\, dp_c\, \langle p_b|2n_0-2j \rangle' \langle p_c|2j
\rangle' J_0 \left( 2\sqrt{n(p_b^2+p_c^2)}\right) \nonumber \\
&=&\frac{1}{2^{2n_0}}\sum_{j=0}^{n_0}\frac{1}{\pi (n_0-j)!j!}
\int dp_b\, dp_c\, H_{2n_0-2j}(p_b) H_{2j}(p_c)
e^{-\frac{p_b^2+p_c^2}{2}} J_0 \left( 2\sqrt{n(p_b^2+p_c^2)}\right)
\nonumber \\
&=&(-1)^{n_0} \int^{\infty}_{0} du\, e^{-u/2} J_0(2\sqrt{nu}) L_{n_0}(u)
\eea
where we changed into polar coordinates $(p,\alpha)$ in the plane
$p_b, p_c$ and applied the identity (see \cite{TRusa})
\be
\int_{0}^{2\pi}d\alpha\, H_{2n}(p\cos\alpha) H_{2m}(p\sin\alpha)
= 2\pi(-1)^{n+m} \frac{(2n)!(2m)!}{n! m!} L_{n+m}(p^2)
\ee
involving Hermite ($H$) and Laguerre ($L$) polynomials. We also
used the normalization condition for the state $|n_0,n_0\rangle$
in the basis $|n_b,n_c\rangle'$ (cf. (\ref{chba})):
\[
\frac{1}{2^{2n_0}}\sum_{j=0}^{n_0}
\frac{(2n_0-2j)!(2j)!}{[(n_0-j)!j!]^2} = 1
\]
With the above expression for $\tilde{f}(n)$ we finally get:
\bea
\int^{\infty}_{0}dn\, \tilde{f}(n)^2&=&
\int^{\infty}_{0} du\, du'\, e^{-\frac{u+u'}{2}}
\int^{\infty}_{0}dn\,
J_0(2\sqrt{nu})J_0(2\sqrt{nu'}) L_{n_0}(u) L_{n_0}(u')\nonumber \\
&=&\int^{\infty}_{0} du\, e^{-u}  L_{n_0}^2(u) ~=~1
\eea
where we have used the orthogonality of Bessel functions and
the normalization of Laguerre polynomials.

\end{document}